\documentstyle[aps,preprint]{revtex}

\begin{document}
\author{H. Aliaga, R. Allub\thanks{%
Member of the Carrera del Investigador Cient\'{\i}fico del Consejo Nacional
de Investigaciones Cient\'{\i}ficas y t\'{e}cnicas (CONICET).} and B. Alascio%
$^{*}$}
\title{Phase diagram for Ca$_{1-x}$Y$_{x}$MnO$_{3}$ type crystals.}
\address{Centro At\'{o}mico Bariloche, (8400) S. C. de Bariloche, Argentina.}
\maketitle

\begin{abstract}
We present a simple model to study the electron doped manganese perovskites.
The model considers the competition between double exchange mechanism for
itinerant electrons and antiferromagnetic superexchange interaction for
localized electrons. It represents each Mn$^{4+}$ ion by a spin 1/2, on
which an electron can be added to produce Mn$^{3+}$; we include a hopping
energy $t$, a strong intratomic interaction exchange $J$ ( in the limit $%
J/t\rightarrow \infty $ ), and an interatomic antiferromagnetic interaction $%
K$ between the local spins.

Using the Renormalized Perturbation Expansion and a Mean Field Approximation
on the hopping terms and on the superexchange interaction we calculate the
free energy. From it, the stability of the antiferromagnetic, canted, ,
ferromagnetic, and novel spin glass phases can be determined as functions of
the parameters characterizing the system.

The model results can be expressed in terms of $t$ and $K$ for each value of
the doping $x$ in phase diagrams.

The magnetization $m$ and canting angle $\theta $ can also be calculated as
fuctions of temperature for fixed values of doping and model parameters.


\end{abstract}


\section{INTRODUCTION}

The discovery of 'colossal ' magnetoresistance (CMR) in La$_{1-x}$Sr$_{x}$MnO%
$_{3}$ type compounds \cite{helm}together with its many unusual properties
have attracted considerable attention. The phase diagram, as a function of
concentration $x$ , temperature, magnetic field, or magnitude of the
superexchange interaction is not quite clear yet for the different compounds.

Before the discovery of CMR, Jonker and Van Santen \cite{jova} established a
temperature-doping phase diagram separating metallic ferromagnetic from
insulating antiferromagnetic phases. Zener \cite{zener} proposed a 'Double
Exchange' (DE) mechanism to understand the properties of these compounds and
the connection between their magnetic and transport properties. This DE
mechanism was used by Anderson and Hasegawa \cite{andha} to calculate the
ferromagnetic interaction between two magnetic ions, and by de Gennes \cite
{degen} to propose canting states for the weakly doped compounds. Kubo and
Ohata \cite{kuo} used a spin wave approach to study the temperature
dependence of the resistivity at temperatures well below the critical
temperature and a mean field approximation at T near T$_{c}.$ Mazzaferro,
Balseiro and Alascio \cite{jorge} used a mixed valence approach similar to
that devised for TmSe combining DE with the effect of doping to propose the
possibility of a metal insulator transition in these compounds.

Since \cite{helm}, a wealth of experimental results have been obtained on
the transport, optical,

spectroscopic and thermal properties of these materials under the effects of
external magnetic fields and pressures \cite{todos}.

Theoretically, Furukawa \cite{furu} has shown that DE is essential to the
transport theory of these phenomena, while Millis {\it et al}. \cite{millis}
have argued that DE alone is not sufficient to describe the properties of
some of the alloys under consideration and have proposed that lattice
polaronic effects play an important role. In a previous paper we have shown
that a semi-phenomenological model, that includes the effect of the disorder
introduced by doping, can explain the transport properties of La$_{1-x}$Sr$%
_{x}$MnO$_{3}$\cite{alal}.

In \cite{alal} we treat the Hamiltonian proposed for these systems using an
alloy analogy approximation to the exchange terms and including the effects
of disorder by introducing a continuous distribution of the diagonal site
energies. Since the focus of the paper was on transport properties of the
ferromagnetic materials, we emphasize the disorder aspects of the problem
and ignore interactions that would give rise to phases other than
ferromagnetic.

Here we propose to extend the previous studies to the region of
concentration of dopant where the antiferromagnetic interactions, always
present, compete with double exchange. We will focus on the Ca rich end of
(RE$)_{x}$Ca$_{1-x}$MnO$_{3}$ compounds (where RE stands for rare earths)
because the antiferromagnetic order is simpler in these alloys.

We include from the start antiferromagnetic interactions between nearest
neighbors, which we treat in mean field approximation. To simplify the
problem we start in this paper by ignoring disorder effects to study the
stability of different phases.

In Section II, we set up the model Hamiltonian and we explain some
approximations which we use in order to solve the model.

Finally, Section III is devoted to obtain the Green's functions relevant to
calculate the kinetic energy of the model, and then we formulate a free
energy that allows us to obtain the different magnetic phases. We discuss
their physical implications and show the results for the density of states
and critical temperatures phase diagrams.

\section{MODEL}

The System in consideration contains two kinds of magnetic ions:

Mn$^{4+}$ with three localized $t_{2g}$ electrons giving rise to a spin 3/2.
We will refer to this electrons as the 'localized spin 'electrons.

Mn$^{3+}$ , which besides the localized spin, contains an itinerant electron
in the $\epsilon _{g}$ orbitals. Due to the strong intra atomic exchange
coupling $J$ , this $\epsilon _{g}$ electron couples ferromagnetically to
the localized spin to produce a spin two at these sites.

The itinerant electron can jump conserving spin from site to site with
hopping energy $t$. These processes give rise to the double exchange
mechanism making the system transport properties metallic like and trying to
order the spins ferromagnetically. It is easy to estimate the energy gain in
the ferromagnetic state due to this process as being of the order of the
hopping energy $t$ times the doping $x$.

Superexchange between localized spins gives rise to an antiferromagnetic
coupling $K$ that competes with double exchange and can lead to different
phases. We investigate here the stability of canted, ferro and
antiferromagnetic phases. To this end, we divide the Mn lattice in two
interpenetrating sublattices appropriate to describe type G
antiferromagnetism \cite{wollan} which is known to be stable at the Ca rich
end of the composition. We indicate by $I$ or $II$ the sites belonging to
each sublattice, and define a quantization axis for each sublattice.

According to the above considerations we write the following model
Hamiltonian:

\begin{equation}
H=\sum_{i,\sigma }\varepsilon _{i}\cdot n_{i,\sigma }+U\sum_{i,\sigma
}n_{i,\sigma }\cdot n_{i,-\sigma }+J\sum_{i}{\bf S}_{i}\cdot {\bf \sigma }%
_{i}+K\sum_{<i,j>}{\bf S}_{i}\cdot {\bf S}_{j}+\sum_{<i,j>,\sigma
}t_{ij}\left( c_{i\sigma }^{+}\cdot c_{j\sigma }+h.c.\right) \,
\end{equation}

\smallskip where $n_{i,\sigma }=${\bf \ }$c_{i\sigma }^{+}$ $c_{i\sigma }$ ,
and $c_{i\sigma }^{+},$ $c_{i\sigma }$ creates and destroys an itinerant
electron with spin $\sigma $ at site $i,$ respectively. {\bf \ }${\bf S}_{i}$
and ${\bf \sigma }_{i}$ are the localized and itinerant spin operators at
site $i$,respectively. $\varepsilon _{i}$ \smallskip are the site diagonal
energies, $U$ is the intra-atomic electronic repulsion, and $t_{ij}$ is the
hopping parameter between nearest neighbors sites $i,j.$

In order to reduce the complexity of the mathematical treatment, we make
from the start two simplifying assumptions that will be discussed later: we
neglect spin flip processes between itinerant and localized electrons ( i.e.
keep only the z component of the ${\bf S}_{i}$ $\cdot $ ${\bf \sigma }_{i}$
interaction) and use a mean field approximation to the antiferromagnetic
interaction ( i.e. we replace ${\bf S}_{i}\cdot {\bf S}_{j}$ by $%
S_{z},_{i}\cdot \langle S_{z,j}\rangle +S_{z},_{j}\cdot \langle
S_{z,i}\rangle $ $-\langle S_{z,j}\rangle .\langle S_{z,i}\rangle $ ).

After doing this, the Hamiltonian can be separated into a local $H_{0}$ and
a non-local $H_{1}.$

\begin{equation}
H=H_{0}+H_{1}  \label{1}
\end{equation}

where:

\begin{equation}
H_{0}=\sum_{i,\sigma }\varepsilon _{i}\cdot n_{i,\sigma }+U\cdot
\sum_{i,\sigma }n_{i,\sigma }\cdot n_{i,-\sigma }+J\cdot
\sum_{i}S_{z,i}\cdot \sigma _{z,i}+K\cdot \sum_{i}S_{z},_{i}\cdot
\sum_{j}\langle S_{z,j}\rangle  \label{2}
\end{equation}

and

\begin{equation}
H_{1}=\sum_{<i,j>,\sigma }t_{ij}\left( c_{i\sigma }^{+}\cdot c_{j\sigma
}+h.c.\right)  \label{3}
\end{equation}

In order to reduce the mathematical complexity of the problem to a minimum
we make the following additional simplifications:

a) We represent the localized spin to be 1/2, so that they can be parallel
or antiparallel to the local quantization direction on each sublattice.

b) We take the semiclassical approximation for the hopping energy: $%
t_{ij}=t\cdot \cos (\theta )$, where $\theta $ is halve the angle between
localized spin directions in $I$ and $II$ sublattices.

c) the diagonal energies $\varepsilon _{i}$ are all equal and define the
zero of one-particle energies. Chemical disorder, which is present in real
samples and fundamental to the transport properties, is ignored in this
first approach. The effect of disorder in the thermodynamic properties is
not crucial to the ordered phases and can be estimated at large disorder by
replacing the hopping energy $t$ by $(t^{2}/\Gamma )$ where $\Gamma $
measures the width of the energy distribution \cite{brial}.

d) Coulomb repulsion $U$ and exchange $\ J$ are much larger than $t_{ij}$
and taken to be infinite here. Finite $U$ and $J$ lead to effective
antiferromagnetic coupling between Mn$^{3+\text{ }}$ions and compete in the
La rich side of the alloys with the ferromagnetic coupling due to double
exchange. The large $U$ limit precludes occupation of the $\epsilon _{g}$
orbitals at each site by more than one itinerant electron.

e) We ignore lattice dynamics that could give rise to polaronic effects.
This phenomena have been discussed by \cite{millis} and references therein.
When necessary, they may be included a posteriori.

\section{RESULTS AND DISCUSSION}

In order to obtain the density of states for itinerant electrons, we have to
calculate local Green's functions. To obtain the local Green's functions, we
resort to the same procedure as in \cite{alal}and \cite{jorge} and work on a
interpenetrating Bethe lattice. The itinerant electrons spin up Green's
functions corresponding to $H_{0}$ at a local spin up site is given by

\begin{equation}
G_{i\uparrow \alpha }^{0l}=<<c_{i\uparrow };c_{i\uparrow }^{\dagger }>>=%
\frac{[\omega -E_{\alpha }-U(1-\bar{n}_{i\downarrow })]}{(\omega -E_{\alpha
})(\omega -E_{\alpha }-U)}\,,
\end{equation}

\smallskip where $\bar{n}_{i\downarrow }=<c_{i\downarrow }^{\dagger
}c_{i\downarrow }>$, $\alpha =+$ ($-$) for up (down) localized spin, $l$ is
the sublattice index ($I,II$), and $E_{\alpha }=\epsilon -\alpha J,$ which
for large $J$ and $U$ reduces to

\smallskip 
\begin{equation}
G_{i\uparrow \alpha }^{0l}=\frac{\delta _{\alpha +}}{\omega -\epsilon +J}\,,
\end{equation}

at the lowest energy pole, which from here on we take at zero ($\epsilon =J)$%
.

The Renormalized Perturbation Expansion (RPE) \cite{eco} as in \cite{alal}
connects the propagator at site $i$ to propagators at the nearest neighbor
sites $i+\delta $ which exclude visiting site $i$ again and which we will
denote by small $g^{\prime }$s. These new propagators are in turn connected
to propagators of the same type at sites $i+\delta +\delta ^{^{\prime
^{\prime }}}$ etc., so that the Green function at each site depends through
this chain on local spin configuration because of the factors $\delta
_{\alpha +}$.

\smallskip The defined chain of equations read for example:

\smallskip 
\begin{equation}
G_{i\uparrow +}^{I}=\frac{1}{\omega -\Delta _{i\uparrow }^{I+}}\,,
\end{equation}

where the self energy $\Delta _{i\uparrow }^{I+}$ is

\smallskip 
\begin{equation}
\Delta _{i\uparrow }^{I+}=\sum_{\delta }\text{ }t_{i,i+\delta }^{2}\text{ }%
g_{i+\delta \uparrow \alpha }^{II}\,,
\end{equation}

in terms of the modified Green's functions $g_{i+\delta \uparrow }^{II}\,$at
the other sublattice, and.

\smallskip 
\begin{equation}
g_{i+\delta \uparrow \alpha }^{II}=\frac{\delta _{\alpha +}}{\omega -\Lambda
_{i+\delta \uparrow }^{II\alpha }}\,,
\end{equation}

\smallskip where $\Lambda $ are the self energies of the new Green's
functions $g,$ which are given in turn by:

\begin{equation}
\Lambda _{i+\delta \uparrow }^{II\alpha }=\sum_{\delta ^{\prime }}\text{ }%
t_{i+\delta ,i+\delta +\delta ^{\prime }}^{2}\text{ }g_{i+\delta +\delta
^{\prime }\uparrow \alpha }^{I}\,,
\end{equation}

and,

\smallskip

\smallskip 
\begin{equation}
g_{i+\delta +\delta ^{\prime }\uparrow \alpha }^{I}=\frac{\delta _{\alpha +}%
}{\omega -\Lambda _{i+\delta +\delta ^{\prime }\uparrow }^{I\alpha }}\,,
\end{equation}

\begin{equation}
\Lambda _{i+\delta +\delta ^{\prime }\uparrow }^{I\alpha }=\sum_{\delta
^{\prime \prime }}\text{ }t_{i+\delta +\delta ^{\prime },i+\delta +\delta
^{\prime }+\delta ^{\prime \prime }}^{2}\text{ }g_{i+\delta +\delta ^{\prime
}+\delta ^{\prime \prime }\uparrow \alpha }^{II}\,,
\end{equation}

etc.

At this point a further approximation is necessary: we take ensemble average
over all possible local spin configurations and we define the magnetization $%
m$ to characterize the background of localized spins at the lattice sites.
Thus, we introduce the probability $\nu _{\alpha }=(1+\alpha \cdot m)/2$
that a site has parallel ( $\alpha =+$) or antiparallel ($\alpha =-$)
localized spin to the quantization axes in each sublattice and we write:

\begin{equation}
G_{\alpha }^{I}=\frac{\nu _{\alpha }}{\omega -\Delta ^{I\alpha }},
\label{15}
\end{equation}

\begin{equation}
\Delta ^{I\alpha }=(k+1)t^{2}(\cos ^{2}(\theta )g_{\alpha }^{II}+\sin
^{2}(\theta )g_{-\alpha }^{II}),  \label{15a}
\end{equation}
\begin{equation}
G_{\alpha }^{II}=\frac{\nu _{\alpha }}{\omega -\Delta ^{II\alpha }},
\label{16}
\end{equation}

\begin{equation}
\Delta ^{II\alpha }=(k+1)t^{2}(\cos ^{2}(\theta )g_{\alpha }^{I}+\sin
^{2}(\theta )g_{-\alpha }^{I}),  \label{16a}
\end{equation}

\begin{equation}
g_{\alpha }^{II}=\frac{\nu _{\alpha }}{\omega -\Lambda ^{II\alpha }},
\label{17}
\end{equation}

\begin{equation}
\Lambda ^{II\alpha }=k\text{ }t^{2}\text{ }(\cos ^{2}(\theta )g_{\alpha
}^{I}+\sin ^{2}(\theta )g_{-\alpha }^{I}),  \label{18}
\end{equation}

\begin{equation}
g_{\alpha }^{I}=\frac{\nu _{\alpha }}{\omega -\Lambda ^{I\alpha }},
\label{21}
\end{equation}

\begin{equation}
\Lambda ^{I\alpha }=k\text{ }t^{2}\text{ }(\cos ^{2}(\theta )g_{\alpha
}^{II}+\sin ^{2}(\theta )g_{-\alpha }^{II}),  \label{22}
\end{equation}

Eliminating the self-energies and using the symmetry between lattices $I$
and $II$ the former equations can be reduced to two interconnected equations:

\begin{equation}
g_{+}=\frac{\nu _{+}}{\omega -k\text{ }t^{2}\text{ }(\cos ^{2}(\theta )\text{
}g_{+}+\sin ^{2}(\theta )\text{ }g_{-})},  \label{25}
\end{equation}

\begin{equation}
g_{-}=\frac{\nu _{-}}{\omega -k\text{ }t^{2}\text{ }(\cos ^{2}(\theta )\text{
}g_{-}+\sin ^{2}(\theta )\text{ }g_{+})},  \label{26}
\end{equation}

from which we finally obtain:

\begin{equation}
G_{\alpha }=\frac{\nu _{\alpha }}{\omega -(k+1)\text{ }t^{2}\text{ }\left(
\cos ^{2}(\theta )\text{ }g_{\alpha }+\sin ^{2}(\theta )\text{ }g_{-\alpha
}\right) },  \label{29}
\end{equation}

where $(k+1)$ is the number of nearest neighbors (six for the simple cubic
Mn lattice).

Solving Eqs. \ref{25} to \ref{29} allow us to obtain the densities of states
per site $\rho (m,\theta ,\omega )=\rho _{+}(m,\theta ,\omega )+\rho
_{-}(m,\theta ,\omega )$, with

\begin{equation}
\rho _{\pm }(m,\theta ,\omega )=%
\mathop{\rm Im}%
(G_{\pm })/\pi .  \label{30}
\end{equation}

For $m=0,$ we obtain the paramagnetic case: $G_{-}=G_{+}$ given by

\begin{equation}
G_{+}=\frac{0.5}{\omega -\frac{(k+1)}{2k}\text{ }\left( \omega +\sqrt{\omega
^{2}-2kt^{2}}\right) }.  \label{31}
\end{equation}

As expected, we see that $G_{\pm \text{ }}$is independent of the canting
angle and the band-width reduces to $(2\sqrt{2kt^{2}}).$

For $m=1,$ the extreme order case occur: $G_{-}=0$ and $G_{+}$ reduces to

\begin{equation}
G_{+}=\frac{1}{\omega -\frac{(k+1)}{2k}\text{ }\left( \omega +\sqrt{\omega
^{2}-4kt^{2}\cos ^{2}(\theta )}\right) },  \label{32}
\end{equation}

this Eq. shows that the maximum band-width occur for the ferromagnetic case $%
(\theta =0)$, this is in agreement with the result of Ref.\cite{furu}. For $%
\theta =\pi /2,$ we obtain the antiferromagnetic case and the density of
states reduces to the delta function.

For $0<m<1$, Eq.\ref{30} shows two different pictures according to the
canting angle $\theta :$

a) for $\pi /2\geq \theta \geq \pi /4,$ we have a density of states with a
central one and two lateral bands as shown in Fig. 1.

b) for $\pi /4\geq \theta \geq 0,$ we have a single band structure similar
to the ferromagnetic case obtained in ref.\cite{alal} and the band-width
decrease with $\theta .$ We show this band structures in Fig. 2.

The density of states allows us to write $x$ as:

\begin{equation}
x=\int_{-\infty }^{\varepsilon _{F}}\rho (m,\theta ,\omega )\text{ }d\omega .
\label{33}
\end{equation}

For a fixed value of doping $x,$ Eq. \ref{33} determine the Fermi energy $%
\varepsilon _{F}.$ Henceforth, we take $k=5$ and the hopping energy $t=1.$

The kinetic energy is given by

\begin{equation}
E_{kin}(m,\theta ,x)=\int_{-\infty }^{\varepsilon _{F}}\rho (m,\theta
,\omega )\text{ }\omega \text{ }d\omega .  \label{34}
\end{equation}

For $x=0.5$, $E_{kin}$ has the lowest energy.

In order to obtain the magnetization and the canted angle as a function of
temperature we need to calculate the free energy. To this purposes, we can
write the following expression for the free energy:

\begin{equation}
F=E_{kin}(m,\theta ,x)+E_{K}(m,\theta ,K)-T\cdot S(m),  \label{35}
\end{equation}

where $E_{K}(m,\theta ,K)$ is the antiferromagnetic superexchange energy
related to the localized spins and given in mean field approximation by:

\begin{equation}
E_{K}(m,\theta ,K)=Km^{2}\cos (2\theta ),  \label{36}
\end{equation}

and $S(m)$ is the entropy term, and we take the simplest possible form
compatible with our earlier approximations:

\begin{equation}
S(m)=\ln (2)-\nu _{+}\ln (2\nu _{+})-\nu _{-}\ln (2\nu _{-}).  \label{37}
\end{equation}

More accurate forms of the entropy valid in the mixed valence regime can be
used, see for example \cite{aligia}.

In Eq.\ref{35} $m$ and $\theta $ take the values corresponding to the
minimum of $F$ for a given $x,$ $K$, and $T$ ($t=1$ is the unit of energy).

At zero temperature, the phase diagram is dominated by the competition
between the DE mechanism and the superexchange energy. For $K<<t$, the
kinetic energy is the most important term in the ground state energy and the
minimum correspond to $m=1$ and $\theta =0$ which define the ferromagnetic
phase (F). For $K>>t$, the relevant term emerges from the superexchange
interactions and the minimum correspond to $m=1$ and $\theta =\pi /2$ which
define the antiferromagnetic phase (AF). When $K$ and $t$ have the same
order, competition between both energies take place and a phase which we
call pseudo spin glass (PSG), defined by $m<1$, appears. We obtain two
different PSG\ phases according to the value of $K/t$. So that, starting
from the F phase and increasing $K$ we obtain first a ferromagnetic pseudo
spin glass (FPSG) characterized by $\theta =0$, and then a second transition
into an antiferromagnetic pseudo spin glass (AFPSG) with $\theta =\pi /2.$
Finally, starting from the AFPSG and increasing $K$ we observe the canted
phase (C), where the minimum corresponds to $m=1$ and $\theta <\pi /2$ (as $%
K $ increases, $\theta \rightarrow \pi /2$). Figures 3 and 4 show this
behaviour for two values of the concentration. PSG phases reduce with $x$
and disappear for $x=0.5$ as shown in Fig. 4.

The presence of the PSG phases, both in the ferro and antiferromagnetic
regimes suggests that the competition between DE and SE interactions gives
rise to frustration rather than to the canted state. This differ from the
results obtained by Arovas and Guinea \cite{Arovas}who used a Schwinger
boson formalism to describe hole doped LaMnO$_{3}$. However, the phase
diagram obtained here, is very similar to that obtained by Golosov et.al.%
\cite{Golosov} . Both, this treatment and ours, allow for local distortions
of the spin arrangement to lower the kinetic energy of itinerant electrons,
we think that the PSG state is a state were local distortions appear in the
ferro or antiferro magnetic phases.

At finite temperature, the phase diagrams can be obtained in a
straightforward way from the free energy $F.$

For small antiferromagnetic interaction ( $K<<t$), starting in the F phase
and increasing the temperature $T$ we find a second order transition into
the paramagnetic phase (P), stable at high temperature and defined by $m=0$
at any value of $\theta $ ( the free energy is independent of $\theta $ ).
For large antiferromagnetic coupling ( $K>>t$), starting in the AF phase and
increasing $T$ we obtain again a second order transition into the
paramagnetic phase (P). In either case, a linear dependence of the critical
temperature ( T$_{c}$) is obtained: T$_{c}\rightarrow \pm $ 2$K$ for $%
K/t\rightarrow $ $\infty $.

When $K$ and $t$ are the same order, the competition between both take place
and different transitions can occur increasing $T$ at low temperatures:

a) Starting from the C phase we obtain a first order transition into AF
phase. We show these transitions in Fig. 3 and Fig. 4.

b) For small values of doping $x$, we obtain a first order transition from
AF into F in a narrow region of the antiferromagnetic coupling. This
transition is depicted in Fig. 3.

c) For $x$ near to 0.5, at very low temperatures the phase diagram shows a
second order transition from C into P. See Fig. 4.

In conclussion, we have studied the competition between double exchange
mechanism for itinerant electrons and antiferromagnetic superexchange
interaction for localized electrons. We approach the problem by truncating
the hamiltonian to reduce the hund energy to a z component coupling and
calculate Green's functions using an RPE and mean field approximation to
obtain the density of states for the itinerant electrons. We have then
calculated the Free energy, to obtain the different phase diagrams for
different dopings. The type of magnetic order assumed makes our calculations
valid only for the electron doped manganese perovskites.

\begin{center}
{\it Acknowledgments}
\end{center}

One of us (R. A.) is supported by the Consejo Nacional de Investigaciones
Cient\'{i}ficas y T\'{e}cnicas (CONICET). B. A. is partially supported by
CONICET.

\vspace{7mm}  

{\large {\bf FIGURE CAPTIONS }} \vspace{3mm}

Figure 1. Total density of states versus energies ($\omega $) for $k/t=5,$ $%
m=0.9,$ and two different values of the canting angle in the region $\pi
/4\leq \theta \leq \pi /2$. The ''side'' bands are related with a process of
hopping electron between a localized spin oriented parallel to its direction
of quantization (+) and a first neighbor localized spin oriented
antiparallel to its corresponding quatization direction (-).

Figure 2. Total density of states as a function of $\omega $ for $k/t=5,$ $%
m=0.9,$ and two different values of the canting angle in the region $0\leq
\theta \leq \pi /4$.

Figure 3. Phase diagram $T/t$ vs $K/t$, with a $x=0.1$ electrons/site.
Different phases appears: paramagnetic (P), ferro (F), antiferro (AF),
canted (C) and pseudo spin glass (FPSG and AFPSG). Transitions into the P
phase are second order. All others transitions are first order.

Figure 4. Phase diagram $T/t$ vs $K/t$, with a $x=0.5$ electrons/site. For
this concentration, the kinetic energy is the lowest and then the PSG phases
dissappear. We can see paramagnetic (P), ferro (F), antiferro (AF), and
canted (C) phases only. Transitions into the P phase are second order. All
others transitions are first order. Note the interphase between F and C
phases.


\begin{references}
\bibitem{helm}  R. von Helmholt, J. Wecker, B. Holzapfel, L. Schultz, and K.
Samwer, Phys. Rev. Lett. {\bf 71}, 2331 (1993).

\bibitem{jova}  G. H. Jonker and J. H. van Santen, Physica {\bf 16}, 337
(1950); J. H. van Santen and G. H. Jonker, Physica {\bf 16}, 599 (1950).

\bibitem{zener}  C. Zener, Phys. Rev. {\bf 82}, 403 (1951).

\bibitem{andha}  P. W. Anderson and H. Hasegawa, Phys. Rev. {\bf 100}, 675
(1955).

\bibitem{degen}  P. G. de Gennes, Phys. Rev. {\bf 118}, 141 (1960).

\bibitem{kuo}  K. Kubo and N. Ohata, J. Phys. Soc. Jpn. {\bf 33}, 21 (1972).

\bibitem{jorge}  J. Mazzaferro, C. A. Balseiro, and B. Alascio, J. Phys.
Chem. Solids {\bf 46}, 1339 (1985).

\bibitem{todos}  Y. Moritomo, A. Asamitsu, and Y. Tokura, Phys. Rev. B {\bf %
51}, 16491 (1995); Y. Okimoto, T. Katsufuji, T. Ishikawa, A. Urushibara, T.
Arima, and Y. Tokura, Phys. Rev. Lett. {\bf 75}, 109 (1995); S. W. Cheong,
H. Y. Hwang, P. G. Radaelli, D. E. Cox, M. Marezio, B. Batlogg, P. Schiffer,
and A. P. Ramirez, Proceedings of the ''Physical Phenomena at High Magnetic
Fields - II'' Conference, Tallahassee, Florida. World Scientific, to be
published;; M. C. Martin, G. Shirane, Y. Endoh, K. Hirota, Y. Moritomo, and
Y. Tokura, To be published; R. Mahendiran, R. Mahesh, A. K. Raichaudhuri,
and C. N. R. Rao, Solid State Commun. {\bf 94}, 515 (1995); H. L. Ju, J.
Gopalakrishnan, J. L. Peng, Qi Li, G. C. Xiong, T. Venkatesan, and R. L. G\`{%
\i }eene, Phys. Rev. B {\bf 51}, 6143 (1995); M. K. Gubkin, A. V. Salesskii,
V. G. Krivenko, T. M. Perekalina, T. A. Khimich, and V. A. Chubarenko, JETP
Lett. {\bf 60}, 57 (1994).

\bibitem{furu}  N. Furukawa, J. Phys. Soc. Jpn. {\bf 63}, 3214 (1994).

\bibitem{millis}  A. J. Millis, P. B. Littlewood, and B. I. Shrainman, Phys.
Rev. Lett. {\bf 74}, 5144 (1995).

\bibitem{alal}  R. Allub and B. Alascio, Solid State Commun. {\bf 99}, 613
(1996); R. Allub and B. Alascio, Phys. Rev. B {\bf 55}, 14113 (1997).

\bibitem{wollan}  E. O. Wollan and W. C. Koehler, Phys. Rev. {\bf 100}, 545
(1955).

\bibitem{brial}  J. Bri\'{a}tico, B. Alascio, R. Allub, A. Butera, A.
Caneiro, M. T. Causa, and M. Tovar, Phys. Rev.B {\bf 53}, 14020 (1996).

\bibitem{muhi}  E. M\"{u}ller-Hartmann and E. Dagotto, to appear in PRB,
cond-mat/9605041.

\bibitem{exp1}  Y. Tokura, A. Urushibara, Y. Moritomo, T. Arima, A.
Asamitsu, G. Kido, and N. Furukawa, J. Phys. Soc. Jpn. {\bf 63}, 3931 (1994).

\bibitem{eco}  See, e.g., E. N. Economou, {\it Green's Functions in Quantum
Physics} Springer Series in Solid-State Sciences {\bf 7}, Ed. P. Fulde.

\bibitem{andy}  P. W. Anderson, Phys. Rev. {\bf 109}, 1492 (1958).

\bibitem{lie}  D. C. Licciardello and E. N. Economou, Phys. Rev. {\bf 11},
3697 (1975).

\bibitem{lloyd}  P. Lloyd, J. Phys. C {\bf 2}, 1717 (1969).

\bibitem{zim}  J. M. Ziman, J. Phys. C {\bf 2}, 1230 (1969).

\bibitem{aligia}  A. A. Aligia Thesis, Instituto Balseiro (1984).

\bibitem{moda}  N. F. Mott and E. A. Davis, {\it Electronic Processes in
Non-Crystallyne Materials}, Oxford University Press (1971).

\bibitem{hwan}  H. Y. Hwang, S.W. Cheong, P. G. Radelli, M. Marezio, and B.
Batlogg, Phys. Rev. Lett.{\bf 75}, 914 (1995).

\bibitem{kuwa}  H. Kuwahara, Y Tomioka, A. Asamitsu, Y. Moritomo, Y. Tokura,
Science {\bf 270}, 961 (1995).

\bibitem{zhao}  Guo-meng Zhao, K. Conder, H. Keller and K. A. Muller, Nature 
{\bf 381}, 676, (1996).

\bibitem{tovar}  J. Bri\'{a}tico, B. Alascio, R. Allub, A. Butera, A.
Caneiro, M. T. Causa, and M. Tovar. Czechoslovak J. Phys. {\bf 46, }S4 2013
(1996).

\bibitem{Arovas}  D. P. Arovas and F. Guinea, preprint

\bibitem{Golosov}  D. I. Golosov, M. R. Norman, and K. Levin,
cond-mat/9712094
\end{references}
\end{document}